\documentclass[pra,twocolumn,showpacs,preprintnumbers]{revtex4}
\usepackage{graphicx}
\usepackage{amsfonts}
\usepackage{amssymb}
\usepackage{amsmath}
\makeindex
\newcommand{\be}{\begin{equation}}
\newcommand{\ee}{\end{equation}}
\newcommand{\bea}{\begin{eqnarray}}
\newcommand{\eea}{\end{eqnarray}}
\newcommand{\Eq}[1]{Eq.\,(\ref{#1})}
\newcommand{\Fig}[1]{Fig.\,\ref{#1}}
\newcommand{\Sec}[1]{Sec.\,\ref{#1}}
\newcommand{\Onlinecite}[1]{Ref.\,\onlinecite{#1}} 

\newcommand{\GFk}{\hat{\mathbf{G}}_k}
\newcommand{\GFkn}{\hat{\mathbf{G}}_{\varkappa_\nu}}

\newcommand{\br}{\mathbf{r}}

\newcommand{\ba}{\mathbf{a}}
\newcommand{\bb}{\mathbf{b}}
\newcommand{\bE}{\mathbf{E}}
\newcommand{\bs}{\boldsymbol{\sigma}}
\newcommand{\En}{\mathbf{E}_n}
\newcommand{\Em}{\mathbf{E}_{n'}}
\newcommand{\Fn}{\mathbf{F}_n}
\newcommand{\Epn}{\mbox{\boldmath ${\cal E}$}\hspace*{-1.5pt}_\nu}

\newcommand{\epshat}{\hat{\boldsymbol{\varepsilon}}}
\newcommand{\heps}{\hat{\boldsymbol{\varepsilon}}}
\newcommand{\leqslantt}{\!\leqslant\!}

\begin{document}
\title{Resonant state expansion applied to three-dimensional open optical systems}
\author{ M.\,B. Doost}
\affiliation{School of Physics and Astronomy, Cardiff University, Cardiff CF24 3AA,
United Kingdom}
\author{W. Langbein}
\affiliation{School of Physics and Astronomy, Cardiff University, Cardiff CF24 3AA,
United Kingdom}
\author{E.\,A. Muljarov}
\affiliation{School of Physics and Astronomy, Cardiff University, Cardiff CF24 3AA,
United Kingdom}
\begin{abstract}
The resonant state expansion (RSE), a rigorous perturbative method in electrodynamics, is developed for three-dimensional open optical systems. Results are presented using the analytically solvable homogeneous dielectric sphere as unperturbed system. Since any perturbation which breaks the spherical symmetry mixes transverse electric (TE) and transverse magnetic (TM) modes, the RSE is extended here to include TM modes and a zero-frequency pole of the Green's function. We demonstrate the validity of the RSE for TM modes by verifying its convergence towards the exact result for a homogeneous perturbation of the sphere. We then apply the RSE to calculate the modes for a selection of perturbations sequentially reducing the remaining symmetry, given by a change of the dielectric constant of half-sphere and quarter-sphere shape. Since no exact solutions are known for these perturbations, we verify the RSE results by comparing them with the results of state of the art finite element method (FEM) and finite difference in time domain (FDTD) solvers. We find that for the selected perturbations, the RSE provides a significantly higher accuracy than the FEM and FDTD for a given computational effort, demonstrating its potential to supersede presently used methods. We furthermore show that in contrast to presently used methods, the RSE is able to determine the perturbation of a selected group of modes by using a limited basis local to these modes, which can further reduce the computational effort by orders of magnitude.

\end{abstract}
\date{\today}
\maketitle
\section{Introduction}

The electromagnetic spectrum of an open optical system is characterized by its resonances, which is evident for optical cavities such as dielectric toroid~\cite{ArmaniN03} or micro-sphere resonators~\cite{CollotEPL93}. Resonances are characterized by their spectral positions and linewidths,  corresponding to, respectively, the real and imaginary part of the complex eigenfrequencies of the system. Finite linewidths of resonances are typical for open systems and are due to energy leakage from the system to the outside. Objects in close proximity of the cavity modify the electromagnetic susceptibility and perturb the cavity resonances, changing both their position and linewidth, most noticeably for the high-quality (i.e. narrow-linewidth) resonances. This effect is the basis for resonant optical biosensors~\cite{VollmerNMe08,LuttiAPL08,ChantadaJOSAB08} in which the changes in the spectral properties of resonators in the presence of perturbations can be used to characterize the size and shape of attached nanoparticles~\cite{ZhuNPho10}. The whispering gallery mode (WGM) resonances in microdisks and spherical microcavities have been used in sensors for the characterization of nanolayers~\cite{NotoOL05}, protein~\cite{VollmerAPL02} and DNA molecules~\cite{VollmerBJ03}, as well as for single atom~\cite{RosenblitPRA04} and nanoparticle detection~\cite{HeNNa11,AltonNP13}.
Furthermore, the long photon lifetime of WGMs can result in their strong coupling to atoms~\cite{VernooyPRA98}.
Recently, optical resonances have become the core element of a more accurate modeling of multimode and random lasers \cite{TureciS08,HischPRL13} and of light propagation through random media~\cite{WangN11}.
In nanoplasmonics, the resonances of metal nanoparticles are used to locally enhance the electromagnetic field~\cite{DanthamNL13}.

Due to the lack of a suited theory, the electromagnetic properties of such open systems were up to now modeled by using finite element method (FEM) and finite difference in time domain (FDTD) solvers. Only recently, approximate approaches using resonance modes have been reported~\cite{TureciPRA06,RubinPRA10,WiersigPRA12,SauvanPRL13,VialPRA14}.  While the eigenmodes of resonators for a few highly symmetric geometries can be calculated exactly, determining the effect of perturbations which break the symmetry presents a significant challenge as the popular computational techniques in electrodynamics, such as the FDTD~\cite{TafloveBook00} or FEM~\cite{WiersigJOA03}, need large computational resources~\cite{BoriskinJOSAA08} to model high quality WGMs.

To treat such perturbations more efficiently, we have developed~\cite{MuljarovEPL10} a rigorous perturbation theory called resonant state expansion (RSE) and applied it to spherical resonators reducible to effective one-dimensional (1D) systems. We have demonstrated on exactly solvable examples in 1D that the RSE is a reliable tool for calculation of wavenumbers and electromagnetic fields of resonant states (RSs)~\cite{DoostPRA12}, as well as transmission and scattering properties of open optical systems. We have recently developed the RSE also for effectively two-dimensional (2D) systems~\cite{DoostPRA13}, and planar waveguides~\cite{LewisARX13}.

In this paper we extend the RSE formulation to arbitrary three-dimensional (3D) open optical systems, compare its performance with FDTD and FEM, and introduce a local perturbation approach. The paper is organized as follows. In \Sec{sec:RSE} we give the general formulation of the RSE for an arbitrary 3D system.  In \Sec{sec:Sphere} we treat the homogeneous dielectric sphere as unperturbed system and introduce the basis for the RSE, which consists of normalized transverse electric (TE) and transverse magnetic (TM) modes and is complemented by longitudinal zero frequency modes. This is followed by examples given in \Sec{sec:application3D} A--C illustrating the method and comparing results with existing analytic solutions, as well as numerical solutions provided by using available commercial software. In \Sec{subsec:local} we demonstrate the performance of the RSE as a local perturbation method for a chosen group of modes by introducing a way to select a suitable subset of basis states. Some details of the general formulation of the method including mode normalization and calculation of the matrix elements are given in Appendices A and B.
\section{Resonant state expansion}\label{sec:RSE}
Resonant states of an open optical system with a local time-independent dielectric susceptibility tensor $\epshat(\br)$ and permeability $\mu=1$ are defined as the eigensolutions of Maxwell's wave equation,
\be
\label{me3D}
\nabla\times\nabla\times\En(\br)=k_n^2\heps(\br)\En(\br)\,,
\ee
satisfying the {\it outgoing wave} boundary conditions. Here, $k_n$ is the wave-vector eigenvalue of the RS numbered by the index $n$, and $\En(\br)$ is its electric field eigenfunction in 3D space. The time-dependent part of the RS wave function is given by $\exp(-i\omega_n t)$ with the complex eigenfrequency $\omega_n=c k_n$, where $c$ is the speed of light in vacuum. As follows from \Eq{me3D} and the divergence theorem, the RSs are orthogonal according to
 \bea
0&=&(k_{n'}^2-k_{n}^2)\int _V d{\bf r}\En(\br)\cdot \heps(\br)\Em(\br)\nonumber\\
&&+\oint _{S_V} dS \left(\En\cdot\frac{\partial\Em}{\partial s}-\Em\cdot\frac{\partial\En}{\partial
s}\right)\,,
 \label{orthog}
 \eea
where the first integral in Eq.\,(\ref{orthog}) is taken over an arbitrary simply connected volume $V$ which includes all system inhomogeneities of $\heps({\bf r})$ while the second integral is taken over the closed surface $S_V$, the boundary of $V$, and contains the gradients $\partial/\partial s$ normal to this surface.

The RSs of an open system form a complete set of functions. This allows us to use RSs for expansion of the Green's function (GF) $\GFk(\br,\br')$ satisfying the same outgoing wave boundary conditions and Maxwell's wave equation with a delta function source term,
\be
- \nabla\times\nabla\times \GFk(\br,\br')+k^2\heps(\br)\GFk(\br,\br')=\hat{\mathbf{1}}\delta(\br-\br')\,,
 \label{GFequ}
\ee
where $\hat{\mathbf{1}}$ is the unit tensor and $k=\omega/c$ is the wave vector of the electromagnetic field in vacuum determined by the frequency $\omega$, which is in general complex. The GF expansion in terms of the direct (dyadic) product of the RS vector fields is given by Ref.\cite{DoostPRA13}
\be \GFk(\br,\br')=\sum _n\frac{\En(\br)\otimes\En(\br')}{2 k(k-k_n)}\,.
\label{ML4}\ee
This expansion requires that the RSs are normalized according to
\bea \label{normaliz}
1+\delta_{k_n,0}&=&\int_V d{\bf r}\En(\br)\cdot\heps(\br)\En(\br)\\
 && +\lim_{k\to k_n}\frac{\displaystyle \oint _{S_V} dS
\left(\En\cdot\frac{\partial\mathbf{E}}{\partial s}-\mathbf{E}\cdot\frac{\partial\En}{\partial
s}\right)}{k^2-k_n^2}\,,
\nonumber
 \eea
where ${\bf E}(k, \br)$ is an analytic continuation of the RS wave function $\En(\br)$ around the point $k_n$ in the complex $k$-plane and $\delta_{k_n,0}$ is the Kronecker delta accounting for a factor of two in the normalization of $k_n=0$ modes.
For any spherical surface $S_R$ of radius $R$, the limit in \Eq{normaliz} can be taken explicitly leading for $k_n\neq0$ modes to
\be
\!1\!\!=\!\!\int_{V_R} \!\!\!\!d{\bf r}\En\cdot\heps\En+\frac{1}{2k^2_n}\oint_{S_R}\!\!\!\! dS \left[\En\!\cdot\!\frac{\partial}{\partial r}r\frac{\partial\En}{\partial r}-r\!\left(\frac{\partial \En}{\partial r}\right)^{\!2}\! \right]
\label{normk}
\ee
where $r=|\br|$, with the origin at the center of the chosen sphere. Static $k_n=0$ modes, if they exist in the GF spectrum, are normalized according to
\be
2=\int\!\!d{\bf r}\En\cdot\heps\En\,.
\label{normk0}
\ee
Their wave functions decay at large distances as $1/r^2$ or quicker, and the volume of integration in \Eq{normaliz} can be extended to the full space for which the surface integral is vanishing. The proofs of Eqs.\,(\ref{normaliz}) and (\ref{normk}) are given in Appendix A.

The completeness of RSs allows us to treat exactly a modified (perturbed) problem
\be
\label{me3Dpert}
\nabla\times\nabla\times\Epn(\br)=\varkappa_\nu^2\bigl[\heps(\br)+\Delta\heps(\br)\bigr]\Epn(\br)\,,
\ee
in which the RS wave vector $\varkappa_\nu$ and the electric field $\Epn$ are modified as compared to $k_n$ and $\En$, respectively, due to a perturbation $\Delta\heps(\br)$ with compact support. We treat this problem by (i) solving \Eq{me3Dpert} with the help of the GF,
\be
\Epn(\br)=-\varkappa_\nu^2 \int d{\bf r}' \GFkn(\br,\br')\Delta\heps(\br')\Epn(\br')\,,
\label{GFsol}
\ee
(ii) using in \Eq{GFsol} the spectral representation \Eq{ML4},
\be\label{GFexpansion} \Epn(\br) = -\varkappa_\nu^2\sum_n\En(\br)\frac{\int d{\bf r}'\En(\br')\cdot\Delta\heps(\br') \Epn(\br') }{2\varkappa_\nu(\varkappa_\nu - k_n)}\,,\ee
and (iii) expanding 
the perturbed wave functions into the unperturbed ones,
\be \Epn(\br) = \sum _n b_{n\nu} \En(\br)\,.
\label{expansion}
\ee
This is the {\it RSE method}. The use of the of the unperturbed GF is an essential element of the RSE as \Eq{GFsol} guarantees that the perturbed wave functions satisfy the outgoing boundary condition. The result of using \Eq{expansion} in \Eq{GFexpansion} is
a linear matrix eigenvalue problem
\be \varkappa_\nu\sum _{n'} (\delta_{nn'}+V_{nn'}/2)b_{n'\nu}=k_n b_{n\nu}\,, \label{RSE1}
\ee
which is reduced, using a substitution $b_{n\nu}=c_{n\nu}\sqrt{\varkappa_\nu/k_n}$\,, to the matrix equation~\cite{MuljarovEPL10}
\be \sum _{n'}
\left(\frac{\delta_{nn'}}{k_n}+\frac{V_{nn'}}{2\sqrt{k_n k_{n'}}}\right)
c_{n'\nu}=\frac{1}{\varkappa_\nu} c_{n\nu}\,. \label{RSE} \ee
This allows us to find the wave vectors $\varkappa_\nu$ and the expansion coefficients $c_{n\nu}$ of the perturbed RSs by diagonalizing a complex symmetric matrix. The matrix elements of the perturbation are given by
\be V_{nn'}=\int \En(\br)\cdot\Delta\heps(\br)\Em(\br)\,d \br\,. \label{Vnm} \ee

In our previous works on RSE~\cite{MuljarovEPL10,DoostPRA13} we derived the intermediate result \Eq{GFexpansion} using Dyson's equation for the perturbed GF. The present way to obtain \Eq{GFexpansion} is equivalent, but is simplifying the treatment by not dealing explicitly with the perturbed GF. We note that in 2D systems the set of RSs of a system is complemented with a continuum of states on the cut of the GF~\cite{DoostPRA13}. In this case, all summations in the above equations include states on the cut which are discretized in numerics to produce a limited subset of isolated poles.


\section{Eigenmodes of a dielectric sphere as basis for the RSE}\label{sec:Sphere}

To apply the RSE to 3D systems we need a known basis of RSs. We choose here the RSs of a dielectric sphere of radius $R$  and refractive index $n_R$, surrounded by vacuum, since they are analytically known.
For any spherically symmetric system, the solutions of Maxwell's equations split into four groups: TE, TM, and  longitudinal electric (LE) and longitudinal magnetic (LM) modes~\cite{StrattonBook41}. TE (TM) modes have no radial components of the electric (magnetic) field, respectively. Longitudinal modes are curl free static modes satisfying Maxwell's wave equation for $k_n=0$. Longitudinal magnetic modes have zero electric field, and since we limit ourself in this work to perturbations in the dielectric susceptibility only, they are not mixed by the perturbation to other types of modes and are thus ignored in the following. Furthermore, owing to the spherical symmetry, the azimuthal index $m$ and longitudinal index $l$ are good quantum numbers of the angular momentum operator and take integer values corresponding to the number of field oscillations around the sphere. For each $l$ value there are $2l+1$ degenerate modes with $m=-l..l$.

Splitting off the time dependence $\propto e^{-i\omega t}$ of the electric fields ${\bf E}$ and ${\bf D}$ and magnetic field ${\bf H}$, the first pair of Maxwell's equations can be written in the form
\be
\nabla\times\bE=i k {\bf H}\,,\ \ \ \ \ \nabla\times{\bf H}=-i k {\bf D}
\label{MEpair1}
\ee
where $k=\omega/c$ and ${\bf D}(\br)=\heps(\br){\bf E}(\br)$. Combining them leads to \Eq{me3D} for the RSs and to \Eq{GFequ} for the corresponding GF. For $k\neq0$ states the second pair of Maxwell's equations,
\be
\nabla\cdot {\bf D}=0\,\ \ \ {\rm and }\ \ \ \nabla\cdot {\bf H}=0\,,
\label{MEpair2}
\ee
is automatically satisfied, since $\nabla\times\nabla=0$. However, if $k=0$, it is not guaranteed that solutions of \Eq{MEpair1} satisfy also \Eq{MEpair2}. The spectrum of the GF given by \Eq{ML4} however includes all modes obeying \Eq{MEpair1}, no matter whether \Eq{MEpair2} is satisfied of not. We find that the LE modes actually do not satisfy \Eq{MEpair2} on the sphere surface, such that Maxwell's boundary condition of continuity of the normal component of ${\bf D}$ across the boundary of the dielectric sphere is not fulfilled. The LE modes are therefore just formal solutions of \Eq{me3D} not corresponding to any physical modes of the system. However, they have to be taken into account for the completeness of the basis used in the RSE.

Following \Onlinecite{StrattonBook41}, the three groups of modes of a homogeneous dielectric sphere can be written as
\bea
{\rm TE:}&\bE=-\br\times \nabla f\ & \mbox{and}\ \ i{\bf H}=\frac{\nabla\times\bE}{k}\,, \nonumber\\
{\rm TM:} & i{\bf H}=-\br\times \nabla f  & \mbox{and}\ \  {\bf E}=\frac{\nabla\times i{\bf H}}{\varepsilon k}\,,
\\
{\rm LE:}&\bE=-\nabla f &\mbox{and}\ \  {\bf H}=0\,, \nonumber
\eea
where $f(\br)$ is a scalar function satisfying the Helmholtz equation
\be
\nabla^2 f+k^2\varepsilon f=0\,,
\label{Helm}
\ee
with the permeability of the dielectric sphere in vacuum given by
\be
\varepsilon(r)= \left\{
\begin{array}{cl}
n_R^2 & {\rm for}\ \  r\leqslant R \\
1 & {\rm for}\ \ r>R\,.\\
\end{array} \right.
\label{eps}
\ee
Owing to the spherical symmetry of the system, the solution of  \Eq{Helm} splits in spherical coordinates \mbox{$\br=(r,\theta,\varphi)$} into the radial and angular components:
\be
f(\br)=R_l(r,k)Y_{lm}(\Omega)\,,
\ee
where $\Omega=(\theta,\varphi)$  with the angle ranges $0\leqslant \theta \leqslant\pi$ and $0\leqslant\varphi\leqslant 2\pi$.  The angular component is given by the spherical harmonics,
\be\label{Spherical_Harmonic} Y_{lm}(\Omega) = \sqrt{\frac{2l+1}{2}\frac{(l-|m|)!}{(l+|m|)!}}P^{|m|}_{l}(\cos\theta)\chi_m(\varphi)\,,  \ee
which are the eigenfunctions of the  angular part of the Laplacian,
\be
\hat{\Lambda}(\Omega) Y_{lm}(\Omega)=-l(l+1)Y_{lm}(\Omega)\,,
\ee
where $P^m_l(x)$ are the associated Legendre polynomials.
Note that the azimuthal functions are defined here as
\be \chi_m(\varphi)=\left\{
\begin{array}{lll}
\pi^{-1/2}\sin(m\varphi) & {\rm for} & m<0\, \\
(2\pi)^{-1/2} & {\rm for} & m=0\, \\
\pi^{-1/2}\cos(m\varphi) & {\rm for} & m>0\,,
\end{array}
\right. \label{chi-n} \ee
in order to satisfy the orthogonality condition without using the complex conjugate, as required by \Eq{orthog}.
The radial components $R_l(r,k)$ satisfy the spherical Bessel equation,
\be
\left [\frac{d^2}{dr^2}+\frac{2}{r}\frac{d}{dr}-\frac{l(l+1)}{r^2}+\varepsilon(r) k^2\right] R_l(r,k) =0
\ee
and have the following form
\be
R_l(r,k)= \left\{
\begin{array}{lll}
j_l(n_R k r)/j_l(n_R k R) & {\rm for} & r\leqslant R\, \\
h_l(kr)/h_l(kR) & {\rm for} & r>R\,, \\
\end{array}
\label{R-analyt} \right. \ee
in which $j_l(z)$ and $h_l(z)\equiv h_l^{(1)}(z)$ are, respectively, the spherical Bessel and Hankel functions of the first kind.

In spherical coordinates, a vector field $\bE(\br)$ can be written as
$$ \bE(r,\theta,\varphi)=E_{r}{\bf e}_r+E_{\theta}{\bf e}_{\theta}+E_{\varphi}{\bf e}_{\varphi}= \begin{pmatrix}
E_{r} \\
E_{\theta}\\
E_{\varphi}\\
\end{pmatrix}\,,
$$
where ${\bf e}_r$, ${\bf e}_{\theta}$, and ${\bf e}_{\varphi}$ are the unit vectors. The electric field of the RSs then has the form

\noindent
\be\mathbf{E}^{\rm TE}_n(\br)=A_l^{\rm TE}R_l(r,k_n)\begin{pmatrix}
0\\[5pt]
\dfrac{1}{\sin\theta}\dfrac{\partial}{\partial\varphi}Y_{lm}(\Omega)\\[10pt]
-\dfrac{\partial}{\partial\theta}Y_{lm}(\Omega)\\
\end{pmatrix} \label{eqn:E_TE} \ee
for TE modes,
\be \mathbf{E}^{\rm TM}_n(\br)=\dfrac{A_l^{\rm TM}(k_n)}{\varepsilon(r) k_n r}\left(
\begin{array}{ccc}
l(l+1)R_l(r,k_n)Y_{lm}(\Omega)\\[5pt]
\dfrac{\partial}{\partial r} r R_l(r,k_n)\dfrac{\partial}{\partial\theta}Y_{lm}(\Omega)\\[10pt]
\dfrac{\partial}{\partial r} \dfrac{r R_l(r,k_n)}{\sin\theta}\dfrac{\partial}{\partial\varphi}Y_{lm}(\Omega)\\
\end{array}
\right) \label{eqn:E_TM} \ee
for TM modes, and
\be \mathbf{E}^{\rm LE}_n(\br)=A^{\rm LE}_l \begin{pmatrix}
\dfrac{\partial}{\partial r} R_l(r,0)Y_{lm}(\Omega) \\[10pt]
\dfrac{R_l(r,0)}{r}\dfrac{\partial}{\partial\theta}Y_{lm}(\Omega)\\[10pt]
\dfrac{R_l(r,0)}{r\sin\theta}\dfrac{\partial}{\partial\varphi}Y_{lm}(\Omega)\\
\end{pmatrix}\,\label{eqn_E_LE} \ee
for LE modes. All the wave functions are normalized according to Eqs.\,(\ref{normaliz})--(\ref{normk0}), leading to the following normalization constants:
\bea
A^{\rm TE}_{l}&=&\sqrt{\frac{2}{l(l+1)R^3(n_R^2-1)}}\,,\nonumber\\
\frac{n_R A^{\rm TE}_{l} }{A^{\rm TM}_{l}(k)}&=&\sqrt{\left[\frac{j_{l-1}(n_R kR)}{j_l(n_R kR)}-\frac{l}{n_R kR}\right]^2+\frac{l(l+1)}{k^2R^2}}\,,\nonumber\\
A^{\rm LE}_{l}&=&\sqrt{\frac{2}{R(n_R^2 l+l+1)}}\,.
\label{A-norm}
\eea

The Maxwell boundary conditions following from \Eq{MEpair1}, namely the continuity of the tangential components of ${\bf E}$ and ${\bf H}$ across the spherical dielectric-vacuum interface, lead to the following secular equations determining the RS wavenumbers $k_n$:
\be
\frac{n_R j_l'(n_R z)}{j_l(n_R z)}-\frac{h_l'(z)}{h_l(z)}=0\,
\label{secularTE}
\ee
for TE modes and
\be
\frac{n_Rj_l'(n_Rz)}{j_l(n_Rz)}-\frac{n^{2}_{R}h_l'(z)}{h_l(z)}-\frac{n^{2}_{R}-1}{z}=0\,\label{secularTM}
\ee
for TM modes, where $z=k_nR$ and $j_l'(z)$ and $h_l'(z)$ are the derivatives of $j_l(z)$ and $h_l(z)$, respectively. While the LE modes are the RSs easiest to calculate due to a simple power-law form of their radial functions,
\be
R_l(r,0)= \left\{
\begin{array}{lll}
(r/R)^l & {\rm for} & r\leqslant R \\
(R/r)^{l+1} & {\rm for} & r>R\,, \\
\end{array}
\label{R0} \right.
\ee
it is convenient to treat them in the RSE as part of the TM family of RSs. Indeed, for $r \leqslant R$ they coincide 
with the TM modes taken in the limit $k_n\to 0$:
\be
\mathbf{E}_n^{\rm LE}(\br)=\sqrt{l(n_R^2-1)} \lim_{k_n\to 0}\mathbf{E}_n^{\rm TM}(\br)\,.
\label{TM2TE}\ee
Note that $k_n=0$ is not a solution of the secular equation (\ref{secularTM}) for TM modes. However, using the analytic dependence of the wave functions of TM modes on $k_n$ [see Eqs.\,(\ref{R-analyt}), (\ref{eqn:E_TM}), and (\ref{A-norm})], the limit \Eq{TM2TE} can be taken in the calculation of the matrix elements containing LE modes. The same limit $k_n\to 0$ has to be carefully approached in the matrix eigenvalue problem \Eq{RSE} of the RSE, as the matrix elements are divergent, due to the $1/\sqrt{k_n}$ factor introduced in the expansion coefficients.
We found that adding a finite negative imaginary part to static poles, $k_nR=-i\delta$, with $\delta$ typically of order $10^{-7}$ (determined by the numerical accuracy) is suited for the numerical results presented in the following section. We have verified this by comparing the results with the ones of the RSE in the form of a generalized linear eigenvalue problem \Eq{RSE1}, which has no such divergence, but its numerical solution is a factor of 2-3 slower in the NAG library implementation.

\section{Application to 3D systems with scalar dielectric susceptibility}\label{sec:application3D}

In this section we discuss the application of the RSE to 3D systems described by a scalar dielectric function $\heps(\br)+\Delta\heps(\br)=\hat{\mathbf{1}} [\varepsilon(r)+\Delta\varepsilon(\br)]$. As unperturbed system we use the homogeneous dielectric sphere of radius $R$ with $\varepsilon(r)$ given by \Eq{eps}, having the analytical modes discussed in \Sec{sec:Sphere}.
We use the refractive index $n_R=2$ of the unperturbed sphere throughout this section  and consider several types of perturbations, namely, a homogeneous perturbation of the whole sphere in \Sec{sec:Hom}, a half-sphere perturbation in \Sec{sec:Halfmoon}, and a quarter-sphere perturbation in \Sec{sec:quartermoon}. We demonstrate in \Sec{subsec:local} the performance of the RSE as a local perturbation method for a chosen group of modes by introducing a way to select a suitable subset of basis states. Explicit forms of the matrix elements used in these calculations are given  in Appendix B.

\begin{figure}[t]
\includegraphics*[width=\columnwidth]{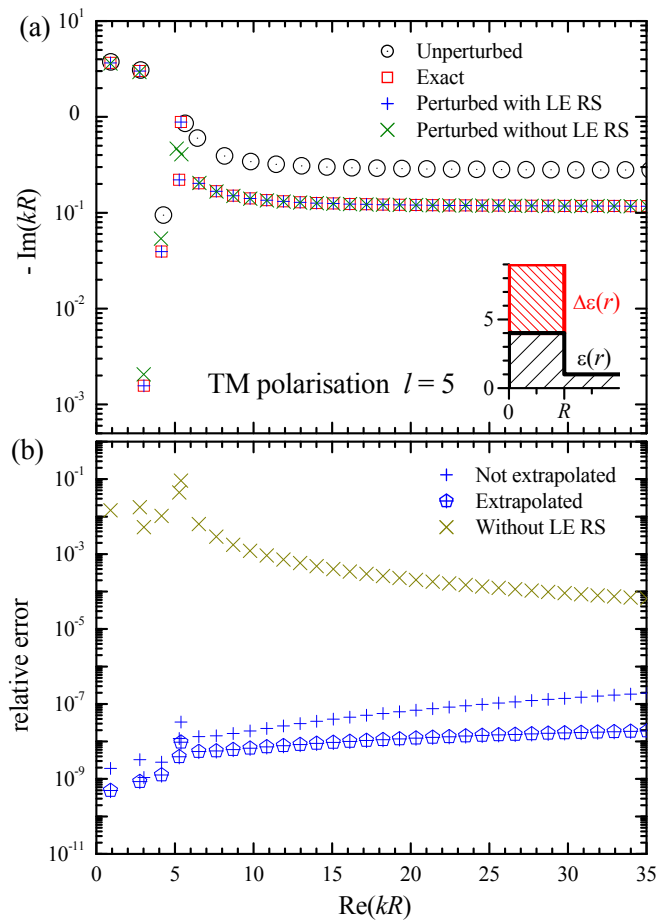}
\caption{TM RSs with $l=5$ (and a fixed $m$) for the homogeneous perturbation \Eq{eps-hom} with $\Delta\epsilon=5$. (a) perturbed RSs wavenumbers calculated using RSE with $N=1000$ with (+) and without ($\times$) the LE mode, as well as using the exact secular equation (open squares). The wavenumbers of the unperturbed system are shown as open circles with dots. Inset: Dielectric constant profile of the unperturbed (black line) and perturbed (red line) systems. (b) Relative error of the perturbed wavenumbers calculated with (+) and without ($\times$) contribution of the LE mode, as well as with the LE mode and extrapolation (crossed heptagons). }\label{fig:F1}
\end{figure}

\begin{figure}[t]
\includegraphics*[width=\columnwidth]{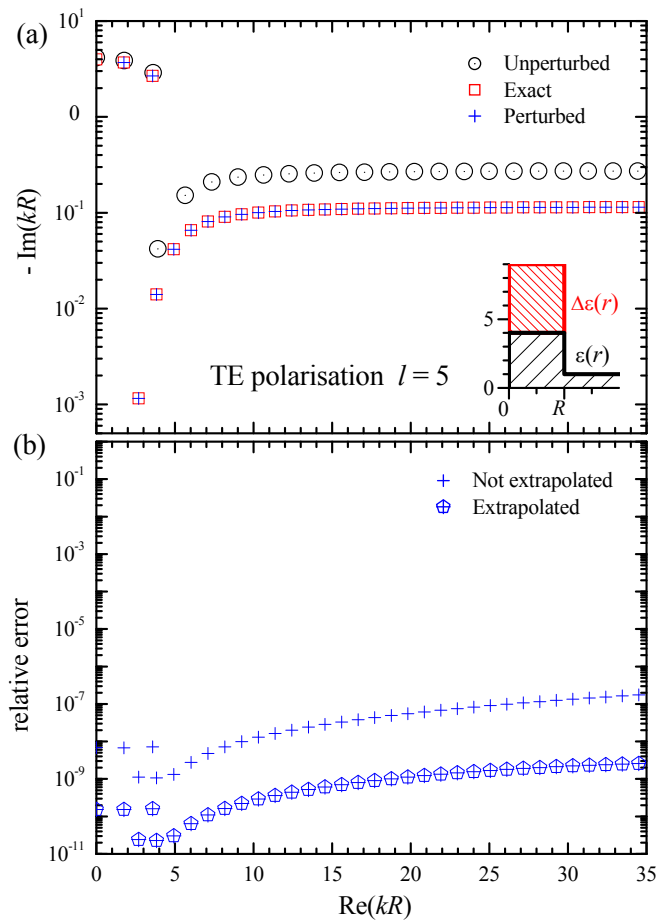}
\caption{As \Fig{fig:F1} but for TE RSs, for which the LE modes have no influence.}\label{fig:F2}
\end{figure}

\subsection{Homogeneous sphere perturbation}
\label{sec:Hom} The perturbation we consider here is a homogeneous change of
$\varepsilon$ over the whole sphere, given by
\be \Delta\varepsilon(\br)=\Delta\epsilon\Theta(R-r)\,,
\label{eps-hom}
\ee
where $\Theta$ is the Heaviside function, with the strength $\Delta\epsilon=5$ used in the numerical calculation. For
spherically symmetric perturbations, RSs of different angular quantum numbers $(l,m)$, and different transverse polarizations are not mixed, and are denenerate in $m$. We show here for illustration the $l=5$ modes. The matrix elements of the perturbation \Eq{eps-hom} are given by Eqs.(\ref{App:TE1})--(\ref{App:F}) of Appendix B. The homogeneous perturbation does not change the symmetry of the system, so that the perturbed modes obey the same secular equations \Eq{secularTE} and \Eq{secularTM} with the refractive index $n_R$ of the sphere changed to $\sqrt{n_R^2+\Delta\epsilon}$, and the perturbed wavenumbers $\varkappa_\nu$ calculated using the RSE can be compared with the exact values $\varkappa^{\rm (exact)}_\nu$ obtained from the secular equations.

We choose the basis of RSs for the RSE in such a way that for a given orbital number $l$ and $m$ we select all RSs with  $|k_n|<k_{\rm max}(N)$ using a maximum wave vector $k_{\rm max}(N)$ chosen to result in $N$ RSs. We find that as we increase $N$, the relative error $\bigl|{\varkappa_\nu}/{\varkappa^{\rm (exact)}_\nu}-1\bigr|$ decreases as $N^{-3}$. Following the procedure described in Ref.\,\onlinecite{DoostPRA12} we can extrapolate the perturbed wavenumbers. The resulting perturbed wavenumbers for $N=1000$ (corresponding to $k_{\rm max}R=800$) are shown in \Fig{fig:F1} for the TM RSs and \Fig{fig:F2} for the TE RSs. The perturbation is strong, leading to WGMs with up to 2 orders of magnitude narrower linewidths. The RSE reproduces the wavenumbers of about 100 RSs to a relative error in the $10^{-7}$ range, which is improving further by one to two orders of magnitude after extrapolation.
The homogeneous perturbation does not couple LE modes to TE modes as LE modes have the symmetry of TM modes [see \Eq{TM2TE}] leading to vanishing overlap integrals with TE RSs.
The contribution of the LE-mode RS in the TM polarization is significant, as is shown in \Fig{fig:F1} by the large decrease of the relative error by up to 8 orders of magnitude when adding them to the basis. This validates the analytical treatment of the LE-mode RSs in the RSE developed in this work. We have verified that taking a finite imaginary value of $\delta=10^{-7}$ in \Eq{RSE} for the LE-modes instead of using strict $k_n=0$ poles in \Eq{RSE1}, as done throughout this work, changes the relative error of the TM mode calculation by less than 10\% and within the range of $10^{-9}$ only. For practical applications, this limitation should not be relevant as the error in the measured geometry will typically be significantly larger.
\subsection{Hemisphere Perturbation}
\begin{figure}
\includegraphics*[width=\columnwidth]{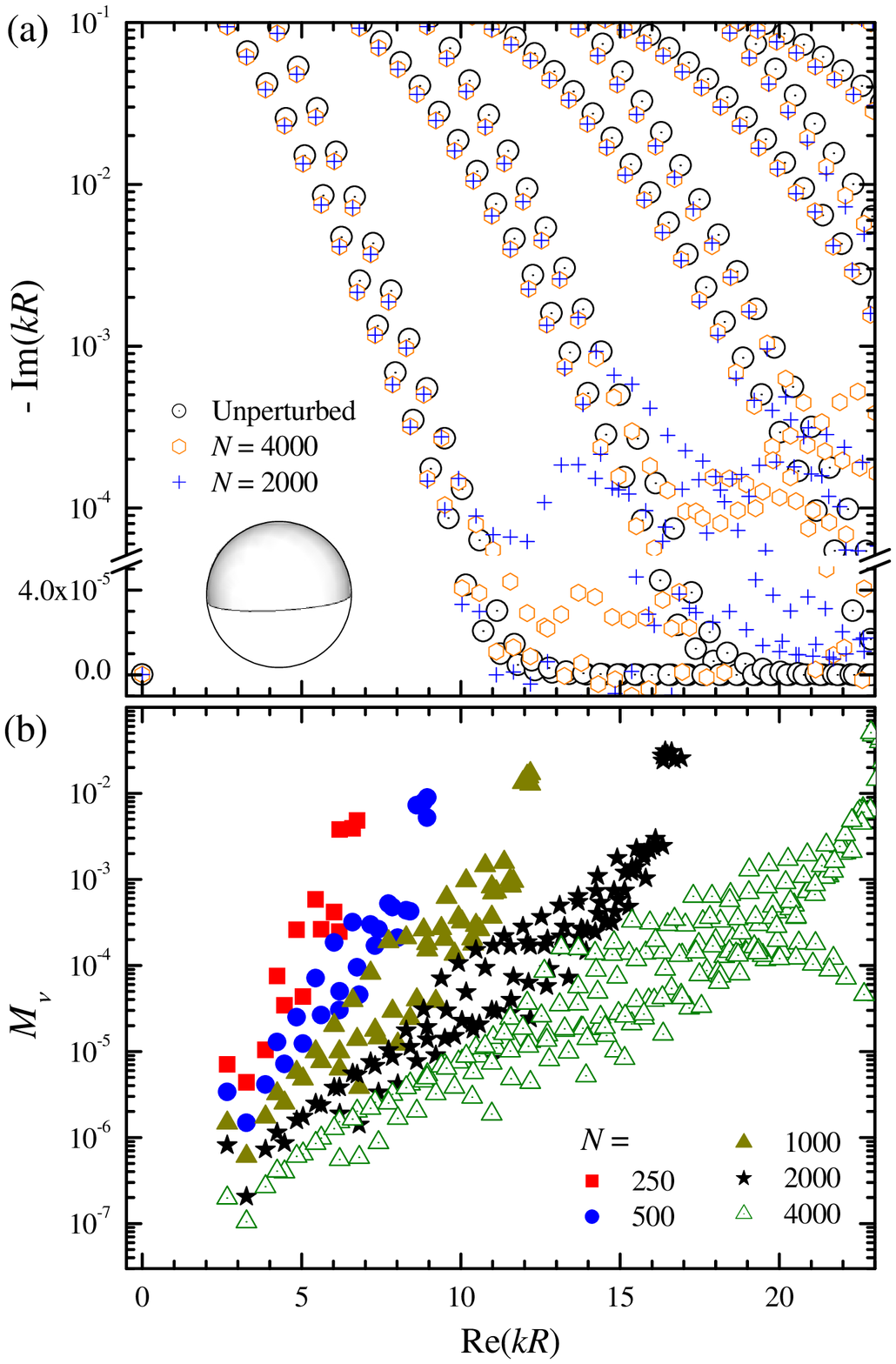}
\caption{
(a) Unperturbed and perturbed RS wavenumbers for a hemisphere perturbation given by \Eq{eps-half} with $\Delta\epsilon=0.2$, for $|m|=3$, calculated via the RSE with basis sizes of $N=2000$ (crosses) and $N=4000$ (hexagons). The unperturbed RSs are shown as open circles with dots. (b) Absolute errors $M_{\nu}$ as function of ${\rm Re}\,\varkappa_{\nu}$ calculated for different basis sizes $N$ as labeled. Inset: Diagram illustrating a dielectric sphere with the regions of increased (lower hemisphere) and decreased (upper hemisphere) dielectric constant.
}\label{fig:F3}
\end{figure}%

\label{sec:Halfmoon} We consider here a hemisphere perturbation as sketched in \Fig{fig:F3} which mixes TE, TM, and LE modes with different $l$, while conserving $m$. The perturbation is given by
\be \Delta\varepsilon(\br)=\Delta\epsilon\Theta(R-r)\Theta\left(\theta-\frac{\pi}{2}\right)
\label{eps-half}\ee
and increases $\varepsilon$ in the northern hemisphere by $\Delta\epsilon$, while leaving the southern hemisphere unchanged. In our numerical simulation, we use $\Delta\epsilon=0.2$. The calculation of the matrix elements is done using Eqs.\,(\ref{App:TETE})--(\ref{App:P}) of Appendix B which require numerical integration. Owing to the symmetry of the perturbation, matrix elements between TM and TE RSs can only be non-zero when the RSs have $m$ of opposite sign and equal magnitude, i.e. they are are sine and cosine states of equal $|m|$. Similarly, matrix elements between two TE RSs or two TM RSs can only be non-zero if both states have the same $m$. We can therefore restrict the basis to $m=3$ TM states and $m=-3$  TE states for the numerical calculations of this section. We treat the LE RSs as TM modes with $k_nR=-i 10^{-7}$ and a normalization factor modified according to \Eq{TM2TE}. The resulting RS wavenumbers are shown in \Fig{fig:F3}. Due to the smaller perturbation compared to that considered in \Sec{sec:Hom}, the mode positions in the spectrum do not change as much. The imaginary part of most of the WGMs decreases due to the higher dielectric constant in the perturbed hemisphere. However, some of the modes also have an increased imaginary part due to the scattering at the edge of the perturbation.

To the best of our knowledge, an analytic solution for this perturbation is not available and thus we cannot calculate the relative error of the RSE result with respect to the exact solution. However, we can investigate the convergence of the method in order to demonstrate how the RSE works in this case, for the perturbation not reducible to an effective one-dimensional problem. We accordingly show in \Fig{fig:F3}(a) the perturbed modes for two different values of basis size $N$ and in \Fig{fig:F3}(b) the absolute errors $M_{\nu}$ for several different values of $N$. Following \Onlinecite{DoostPRA12}, the absolute error is defined here as $M_{\nu}=\max_{i=1,2,3}|\varkappa_\nu^{N_4}-\varkappa_\nu^{N_i}|$, where $\varkappa_\nu^{N_i}$ are the RS wavenumbers calculated for basis sizes of $N_1\approx N/2$, $N_2\approx N/\sqrt{2}$,
$N_3\approx N/\sqrt[4]{2}$, and $N_4=N$. We see that the perturbed resonances are converging with increasing basis size, approximately following a power law with an exponent between $-2$ and $-3$.

\subsection{Quarter-Sphere Perturbation}
\begin{figure}
\includegraphics*[width=\columnwidth]{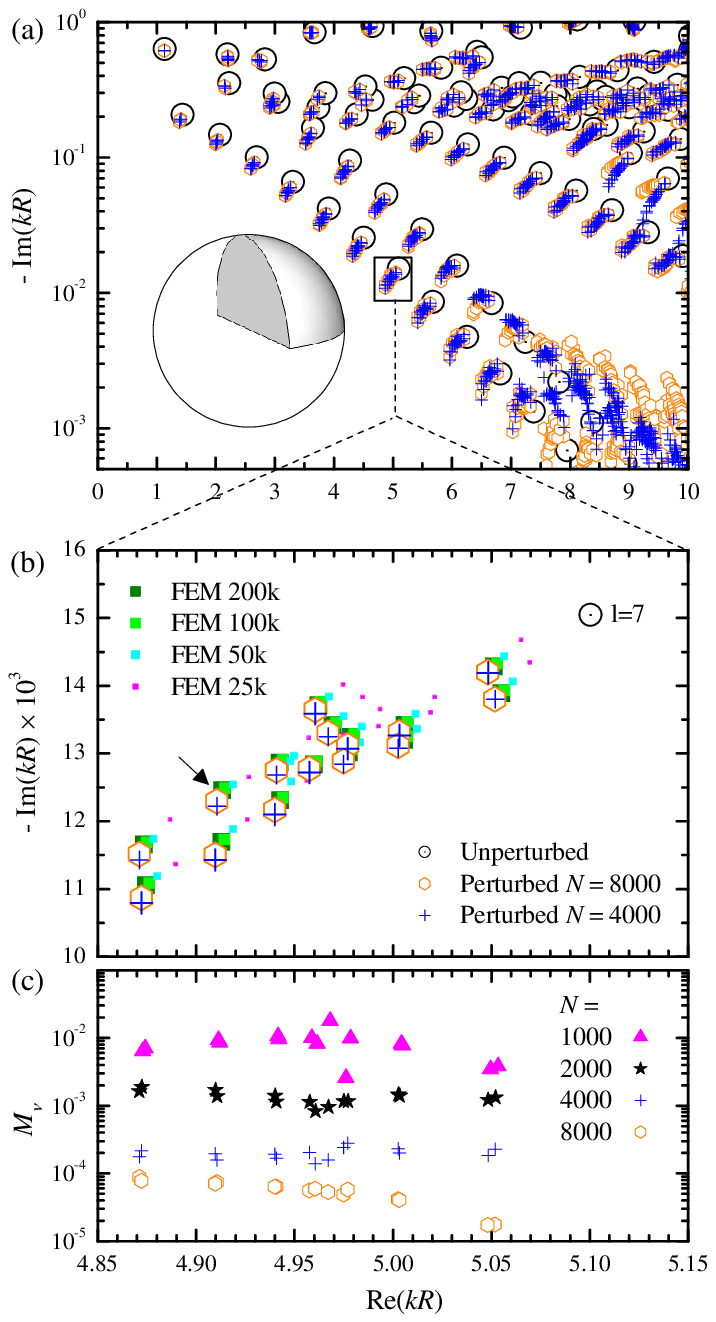}
\caption{(a) Unperturbed and perturbed RS wavenumbers for a quarter-sphere perturbation given by \Eq{eps-Quarter} with $\Delta\epsilon=1$, calculated by the RSE with the basis sizes $N=4000$ (crosses) and $N=8000$ (hexagons). The unperturbed RSs are shown as open circles with dots. A sketch of the perturbation geometry is also shown.  (b) Zoom of (a) showing the splitting of a $2l+1$ degenerate WGMs as the $m$ degeneracy is lifted. Here $l=7$. The pole indicated in (b) by an arrow is analyzed further in \Fig{fig:F6}.  The results of FEM simulations using 200k, 100k, 50k and 25k finite elements are shown for comparison. (c) Absolute error $M_{\nu}$ as function of ${\rm Re}\,\varkappa_{\nu}$ calculated by the RSE with different basis sizes $N$ as labeled, for the RSs shown in (b). }\label{fig:F4}
\end{figure}
\label{sec:quartermoon}
We consider here a perturbation which breaks both continuous rotation symmetries of the sphere and is thus is not reducible to an effective one or two-dimensional system. The perturbation is given by
\be \Delta\varepsilon(\br) = \Delta\epsilon\Theta(R-r)\Theta\left(\frac{\pi}{2}-\theta\right)\Theta\left(\frac{\pi}{2}-|\varphi-\pi|\right)
\label{eps-Quarter}
\ee
and corresponds physically to a uniform increase of the dielectric constant in a quarter-sphere area, as sketched in \Fig{fig:F4}. In our numerical simulation, we take $\Delta\epsilon=1$.
Again, the calculation of the matrix elements requires numerical integration. Owing to the reduced symmetry of the perturbation as compared to that treated in the previous section we now have modes of different $l$, $m$, and polarization mixing, although TE sine (TM cosine) and TE cosine (TM sine) modes are decoupled, owing to the mirror symmetry of the system. This allows us to split the simulation of all modes into two separate simulations called A and B, respectively, each of size $N$. The lifting of the $m$-degeneracy of the unperturbed modes can be seen as splitting off resonances in \Fig{fig:F4}(a) and (b). In most cases the splitting in the real part of the resonant wavenumber is greater than the linewidth of the modes.

The convergence of the RSE is well seen in \Fig{fig:F4}(a) and (b) showing the perturbed RS wavenumbers for two different basis sizes $N$. An analytic solution for this perturbation is not available, so that we use the method described in \Sec{sec:Halfmoon} to estimate the error, and show in \Fig{fig:F4}(c) the resulting absolute errors $M_{\nu}$ for several values of $N$. A convergence with a power law exponent between $-2$ and $-3$ is again observed, resulting in relative errors in the $10^{-4}$ to $10^{-5}$ range for $N=8000$.

To verify the RSE results, we have simulated the system using the commercial solver ComSol (http://www.comsol.com) which uses the finite element method and Galerkin's method, approximating the openness of the system with an absorbing perfectly matched layer (PML). We have surrounded the sphere with a vacuum shell followed by a PML shell of equal thickness $D$. The results are shown in \Fig{fig:F4}(b) using $D=R/2$, and a ``physics controlled'' mesh with $N_{\rm G}=25$k, 50k, 100k and 200k finite elements. We used the nearest unperturbed RS wave vector as linearization point (i.e. the input value) for the ComSol solver, and requested the determination of 40 eigenfrequencies, which we found to be the minimum number reliably returning all 15 non-degenerate modes deriving from the $l=7$ unperturbed fundamental WGM. With increasing $N_{\rm G}$, the ComSol RS wavenumbers tend towards the RSE poles, with an error scaling approximately as $N_{\rm G}^{-1}$. This is verifying the validity of the RSE results.

\begin{figure}
\includegraphics*[width=\columnwidth]{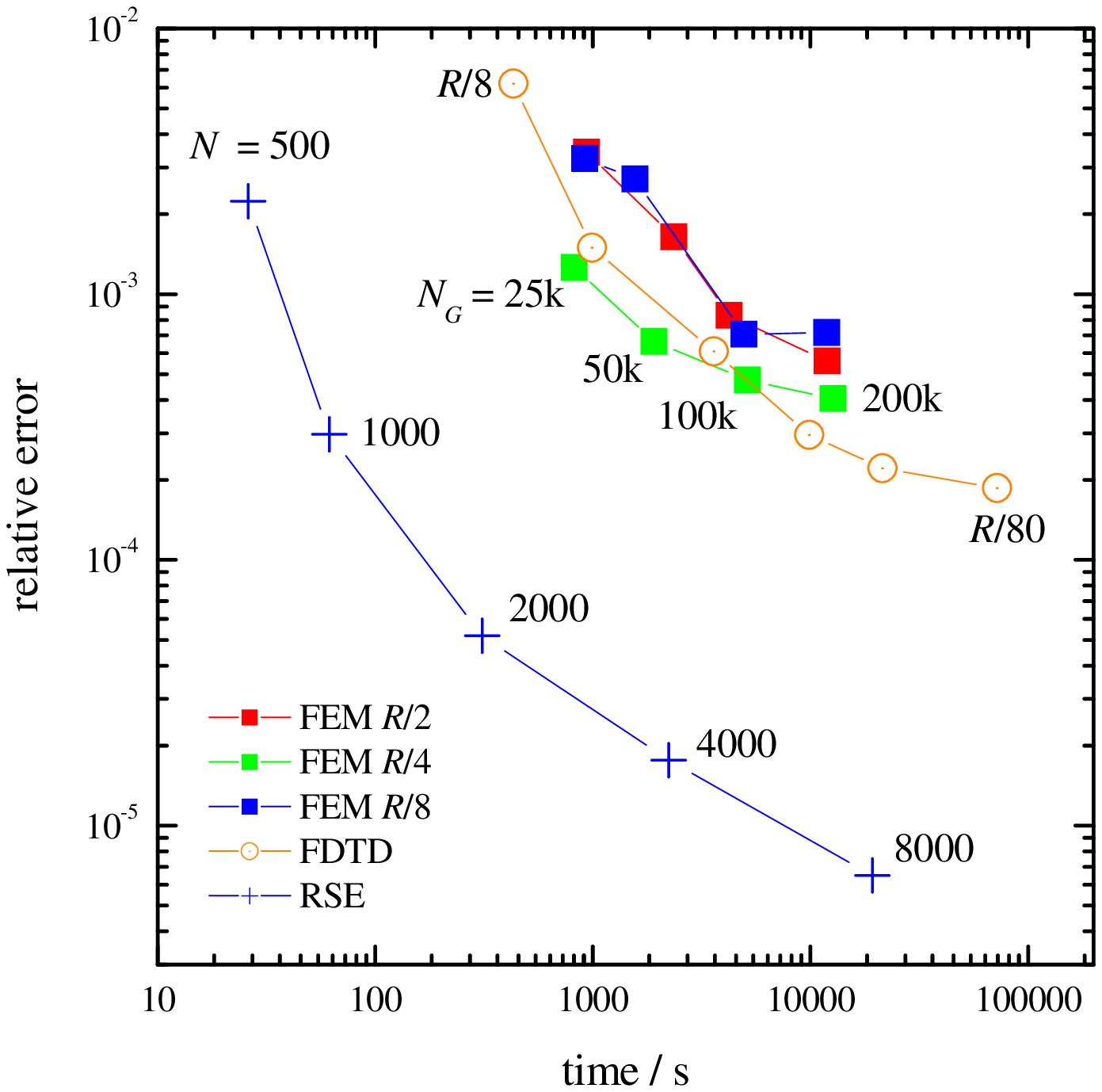}
\caption{A comparison of the relative error of the perturbed RS
wavenumbers shown in \Fig{fig:F4}(b) calculated by the RSE for different $N$ as labeled versus computational time. For comparison, the performance of the FEM using ComSol, and FDTD using Lumerical are given. In the FEM we have used a thickness of the vacuum layer and the perfectly matched layer of $R/2$, $R/4$, and $R/8$ as labeled, and $N_{\rm G}$=25k, 50k, 100k, 200k finite elements as labeled. In the FDTD we used different grid spacings from $R/8$ to $R/80$ and other parameters as given in the text.}\label{fig:F5}
\end{figure}

To make a comparison between the RSE and ComSol in terms of numerical complexity  we use the poles computed by an  $N=16000$ RSE simulation as ``exact solution'' to calculate the average relative errors of the poles shown in \Fig{fig:F4}(b) versus effective processing time on an Intel E8500 CPU. The result is shown in \Fig{fig:F5}, including ComSol data for different shell thicknesses $D$ of $R/2$, $R/4$, and $R/8$, revealing that $D=R/4$ provides the best performance.
This comparison shows that the RSE is 2-3 orders of magnitude faster than ComSol for the present example, and at the same time determines significantly more RSs.

The RSE computing time includes the calculation of the matrix elements which were done evaluating the 1-dimensional integrals (see Appendix \ref{App:Arbpert}) using $10000$ equidistant grid points. The computing time of the matrix elements is significant only for $N\lesssim 2000$, while for larger $N$ the matrix diagonalization time, scaling as $N^3$, is dominating. We have verified that the accuracy of the matrix element calculation is sufficient to not influence the relative errors shown.

We also include in \Fig{fig:F5} the performance of FDTD calculations using the commercial software Lumerical (http://www.lumerical.com). They were undertaken using a simulation cube size from 2.5$R$ to 4$R$, exploiting the reflection symmetry, and for grid steps between $R/8$ and $R/80$, with a sub-sampling of 32. The simulation area was surrounded by a PML of a size chosen automatically by the software. The excitation pulse had a center wavenumber of $kR=5.1$ and a relative bandwidth of 10\% to excite the relevant modes, and the simulation was run for 360 oscillation periods. The calculated time-dependent electric field after the excitation pulse was transformed into a spectrum and the peaks were fitted with a Lorentzian to determine the real and imaginary part of the mode. The parameters used were chosen to optimize the performance, and in the plot the results with the shortest computation time for a given relative error are given.

We can conclude that the RSE is about two orders of magnitude faster than both FEM and FDTD for this specific problem, showing its potential to supersede presently used methods. A general analysis of the performance of RSE relative to FEM and FDTD is beyond the scope of this work and will be presented elsewhere.

\begin{figure}
\includegraphics*[width=\columnwidth]{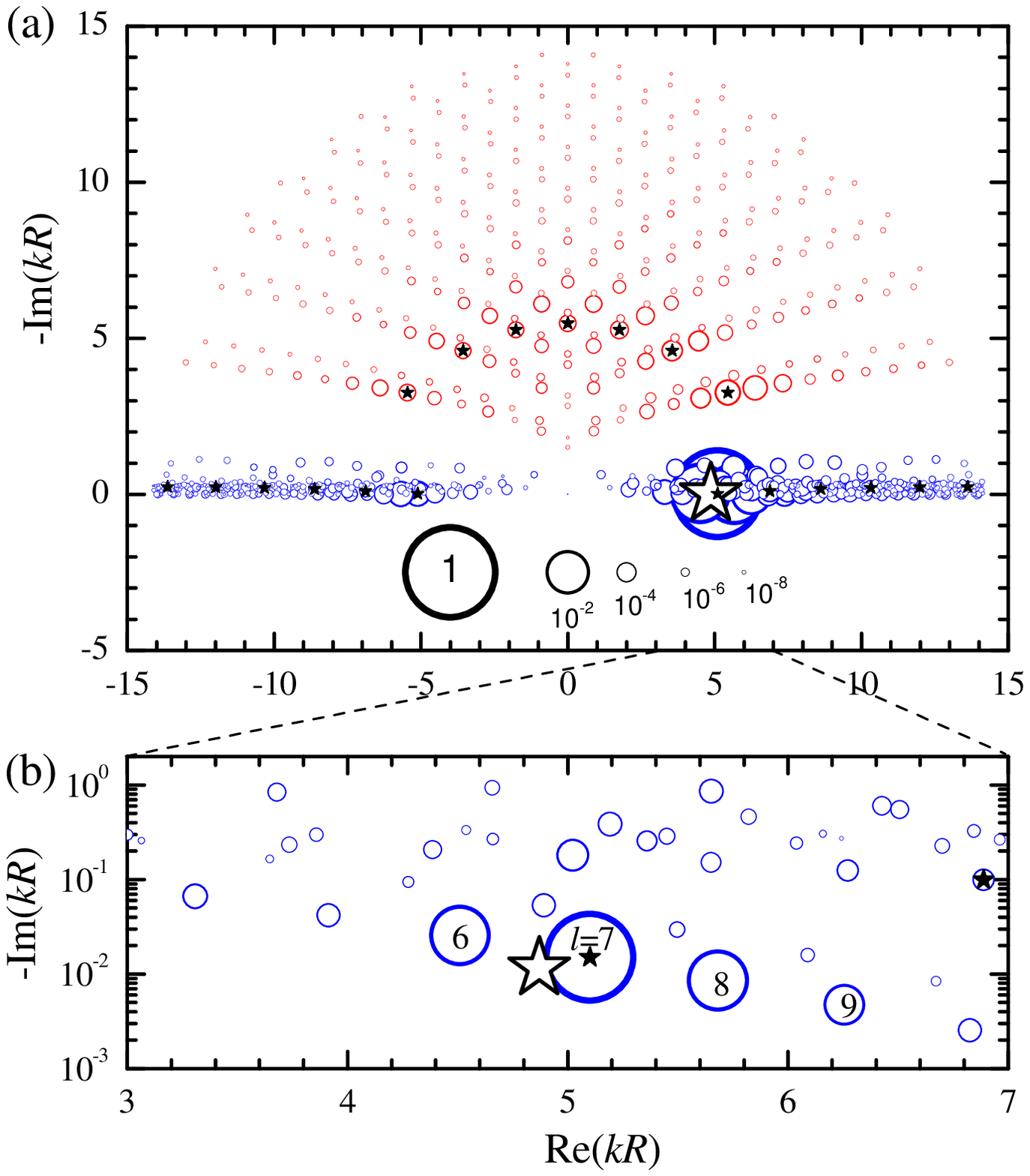}
\caption{(a) Contributions of the basis RSs (blue and red circles) 
to the perturbed RS (open star) indicated by an arrow in \Fig{fig:F5}(b), calculated using the RSE with $N=8000$. Small stars show the positions of $l=7$ TE modes. All circles and stars are centred at the positions of the corresponding RS wavenumbers in the complex $k$-plane.  The radius of the circles is proportional to $\sqrt[6]{\sum |c_{n\nu}|^2}$, where the sum is taken over all $m$-degenerate RSs of the basis system corresponding to the given eigenfrequency. A key showing the relationship between circle radius and $\sum|c_{n\nu}|^{2}$ is given as black circles. (b) A zoom of (a) showing the contribution of RSs close to the chosen perturbed state. The angular quantum numbers $l$ of the WGMs with the largest contributions are indicated.}\label{fig:F6}
\end{figure}

To illustrate how a particular perturbed RS is created as a superposition of unperturbed RSs, we show in \Fig{fig:F6} the contributions of the unperturbed RSs to the perturbed WGM indicated by the arrow in \Fig{fig:F4}(b) with index $\nu$ and wavenumber $\varkappa_\nu$, given by the open star in \Fig{fig:F6}.
The contribution of the basis states to this mode are visualized by circles of a radius proportional to $\sqrt[6]{\sum |c_{n\nu}|^2}$, where the sum is taken over the $2l+1$ degenerate basis RSs of a given eigenfrequency, centered at the positions of the RS wavenumbers in the complex $k$-plane. The expansion coefficients $c_{n\nu}$ decrease quickly with the distance between the unperturbed and perturbed RS wavenumbers, with the dominant contribution coming from the nearest unperturbed RS, a typical feature of perturbation theory in closed systems. The unperturbed RS nearest to the perturbed one in \Fig{fig:F6} has the largest contribution, and is a $l=7$ TE WGM with the lowest radial quantum number. Other WGMs giving significant contributions have the same radial quantum number and the angular quantum numbers ranging  between $l=6$ and $l=9$, see the small stars in \Fig{fig:F6} corresponding to $l=7$ basis states. This is a manifestation of a quasi-conservation of the angular momentum $l$ for bulky perturbations like the quarter-sphere perturbation considered here. 

Generally, we see that a significant number of unperturbed RSs are contributing to the perturbed RS, which is indicating that previous perturbation theories for open systems would yield large errors for the strong perturbations treated in this work since they are limited to low orders~\cite{LeungPRA94,TeraokaJOSAB03} or to degenerate modes only~\cite{LaiPRA90}.

\subsection{Local Perturbation}
\label{subsec:local}
The weights of the RSs shown in \Fig{fig:F6} indicate that a perturbed mode can be approximately described by a subset of the unperturbed modes, which typically have wavenumbers in close proximity to that of the perturbed mode. It is therefore expected that a local perturbation approach based on the RSE is possible. We formulate here such an approach.

We commence with a small subset $\cal{S}$ of modes of the unperturbed system which are of particular interest, for example because they are used for sensing. To calculate the perturbation of these modes approximately, we consider a global basis $\cal{B}$ as used in the previous sections, with a size $N$ providing a sufficiently small relative error.
We then choose a subset $\cal{S}^+\!\! \subset \cal{B}$ with $N'<N$ elements containing $\cal{S}$, i.e. $\cal{S}\!\!\subset \cal{S}^+$, and solve the RSE  \Eq{RSE} restricted to $\cal{S}^+$. The important step in this approach is to find a numerically efficient method to choose the additional modes in $\cal{S}^+$ which provide the smallest relative error of the perturbed states deriving from $\cal{S}$ for a given $N'$. Specifically, the method should be significantly faster than the matrix diagonalization \Eq{RSE}.

To develop such a method, we consider here the Rayleigh-Schr\"odinger perturbation theory based on the RSE and expand the RS wave vector $\varkappa$ up to second order,
\be
\frac{1}{\varkappa}=\left(\frac{1}{\varkappa}\right)^{\!\!\!(0)}\!\!\!+\left(\frac{1}{\varkappa}\right)^{\!\!\!(1)}
\!\!\!+\left(\frac{1}{\varkappa}\right)^{\!\!\!(2)}+\dots\,,
\ee
where
\be
\left(\frac{1}{\varkappa}\right)^{\!\!\!(0)}\!\!\!=\frac{1}{k_n}\,,\ \ \left(\frac{1}{\varkappa}\right)^{\!\!\!(1)}\!\!\!=\frac{V_{nn}}{2k_n}\,,\ \ \left(\frac{1}{\varkappa}\right)^{\!\!\!(2)}\!\!\!=-\frac{1}{4}\sum_{n'\neq n} \frac{V_{nn'}^2}{k_n-k_{n'}}
\label{Second}
\ee
as directly follows from \Eq{RSE}. Note that the second-order result in \Eq{Second} is different from that given in \Onlinecite{LeungPRA94}.

We expect that the second-order correction given by \Eq{Second} is a suited candidate to estimate the importance of modes.
We therefore sort the modes in ${\cal B}$ according to the weight $W_n$ given by
\be
W_n=\sum_{n'\in \cal{D}}\sum_{n''\in \cal{S}}\left|\dfrac{V_{n'n''}^2}{k_{n'}-k_{n''}}\right|,
\label{eqn:Wn}\ee
where $\cal{D}$ is the set of modes degenerate with the mode $n$ in $\cal{B}$. The summation over all degenerate modes is motivated by their comparable contribution to the perturbed mode, as known from degenerate perturbation theory. We add modes of $\cal{B}$ to $\cal{S}^+$ in decreasing $W_n$ order. Groups of degenerate modes $\cal{D}$ are added in one step as they have equal $W_n$. A special case are the LE modes in the basis of the dielectric sphere, which are all degenerate having $k_n=0$. They are added in groups of equal $l$ in the order of reducing weight.

\begin{figure}
\includegraphics*[width=\columnwidth]{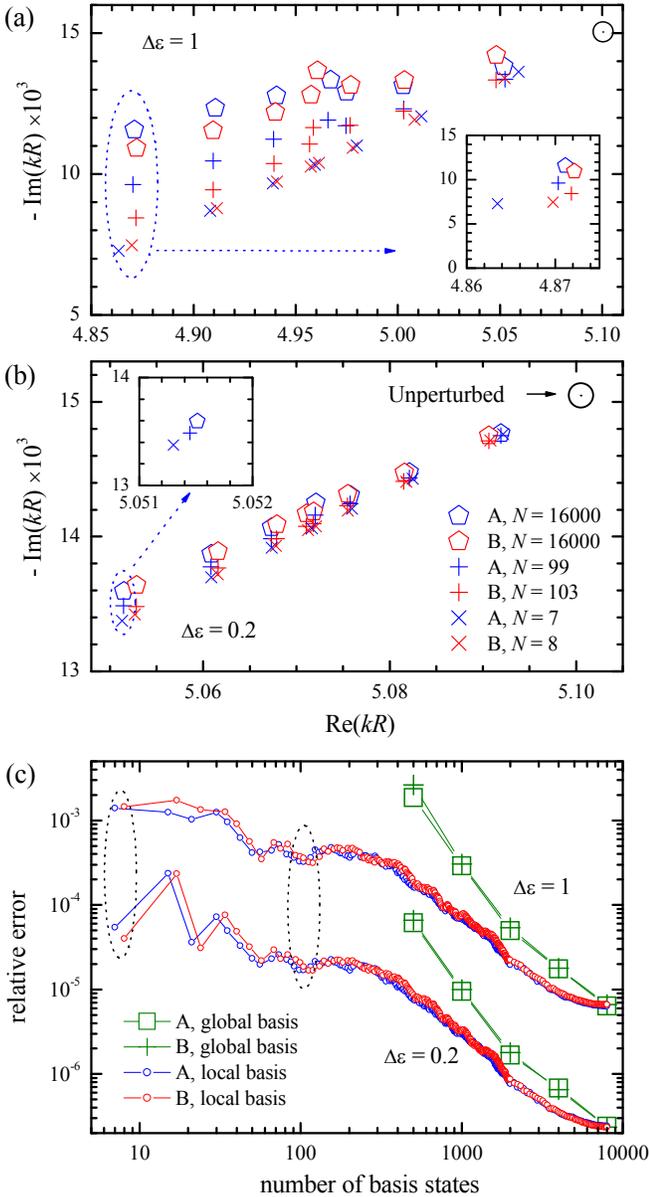}
\caption{(a) Unperturbed and perturbed RS wavenumbers for a quarter-sphere perturbation given by \Eq{eps-Quarter} with $\Delta\epsilon=1$, calculated by the RSE using the local basis sizes $N'=7,8$ (+), $N'=99,103$ ($\times$) for the parts A,B, respectively,  and a global basis with $N=16000$ (hexagons). The unperturbed RSs are shown as a circle with a dot. The inset is a zoom to the RS with the strongest perturbation.  (b) As in (a) but for $\Delta\epsilon=0.2$. (c) Average relative error of the states shown in (a) and (b) versus basis size for a global basis (squares and crosses), and for a local basis (circles) derived from a global basis of $N=8000$ modes.
}\label{fig:F7}
\end{figure}

To exemplify the local perturbation method, we use the quarter sphere perturbation with two different perturbations strengths $\Delta\epsilon=1$ and $\Delta\epsilon=0.2$, and choose the degenerate $l=7$ modes shown in \Fig{fig:F4}(b) as $\cal{S}$. The perturbed RSs deriving from $\cal{S}$ are shown in \Fig{fig:F7}(a) and (b), as calculated by RSE using either a global basis ${\cal B}$ with $N=16000$, or a minimum local basis $\cal{S}^+=\cal{S}$  with $N'\sim10$, or a larger $\cal{S}^+$ with $N'\sim 100$. As in the previous section we show the results separately for each class of RSs (A and B) decoupled by symmetry. We find that for $\Delta\epsilon=0.2$ ($\Delta\epsilon=1$) the perturbation lifts the degeneracy of $\cal{S}$ by a relative wavenumber change of about 1\% (5\%), and that the minimum local basis $\cal{S}^+=\cal{S}$ of only degenerate modes reproduces the wavenumbers with a relative error of about $10^{-4}$ ($10^{-3}$), i.e. the perturbation effect is reproduced with an error of a few \%. Increasing the local basis size to $N'\sim 100$ the error reduces by a factor of three, by similar absolute amounts in the real and the imaginary part of the wavenumber [see insets of \Fig{fig:F7}(a) and (b)].

The relative error of the local-basis RSE is generally decreasing with increasing basis size, as shown in \Fig{fig:F7}(c). It can however be non-monotonous on the scale of individual sets of degenerate modes. This is clearly seen for for $\Delta\epsilon=0.2$ and small $N'$, where adding the second group increases the error, which is reverted when the third group is added. These groups are the $l=6$ and $l=8$ fundamental WGMs as expected from \Fig{fig:F6}(b), which are on opposite sides of $S$ ($l=7$ WGMs) in the complex frequency plane. Adding only one of them therefore imbalances the result, leading to an increase of the relative error.

Comparing results in \Fig{fig:F7}(c) for two different values of $\Delta\epsilon$, we see that the second-order correction dominates the relative error, as in the wide range of $N'$ the error scales approximately like a square of the perturbation strength. The global-basis RSE, also shown in \Fig{fig:F7}(b), has for a given basis size significantly larger errors. Furthermore, a minimum basis size is required for the basis to actually contain $\cal{S}$, in the present case $N\approx 500$. The local basis thus provides a method to calculate the perturbation of arbitrary modes with a small basis size.

The local perturbation method described in this section enables the calculation of high frequency perturbed modes which have previously been numerically inaccessible to  FDTD and FEM due to the necessity of the corresponding high number of elements needed to resolve the short wavelengths involved and inaccessible to the RSE with a global basis due to the prohibitively large $N$ required. The example we used for the illustration shows that a basis of $\sim 100$ RSs in the local RSE can be sufficient to achieve the same accuracy as provided by FDTD and FEM in a reasonable computational time [see Figs.\,\ref{fig:F5} and \ref{fig:F7}(c)]. For this basis size, solving the RSE \Eq{RSE} is 6 orders of magnitude faster than FDTD and FEM, and the computational time in our numerical implementation is dominated by the matrix element calculation which can be further optimized. A detailed evaluation of the performance of the local basis RSE and a comparison of selection criteria different from \Eq{eqn:Wn} will be given in a forthcoming work.

\section{Summary}\label{sec:summary}
We have applied the resonant state expansion (RSE) to general three-dimensional (3D) open optical systems. This required including in the basis both types of transversal polarization states, TE and TM modes, as well as longitudinal electric field modes at zero frequency. Furthermore, a general proof of the mode normalization used in the RSE is given. Using the analytically known basis of resonant states (RSs) of a dielectric sphere -- a complete set of eigenmodes satisfying outgoing wave boundary conditions -- we have applied the RSE to perturbations of full-, half- and quarter-sphere shapes. The latter does not have any rotational or translational symmetry and is thus not reducible to lower dimensions, so that their treatment demonstrates the applicability of the RSE to general 3D perturbations.

We have compared the performance of the RSE with commercially available solvers, using both the finite element method (FEM) and finite difference in time domain (FDTD), and showed that for the geometries considered here, the RSE is several orders of magnitude more computationally efficient, showing its potential to supersede presently used computational methods in electrodynamics. We have furthermore introduced a local perturbation method for the RSE, which is restricting the basis in order to treat a small subset of modes of interest.  This further reduces computational efforts and improves on previous local perturbation methods.

\acknowledgments M.\,D. acknowledges support by the EPSRC under the DTA scheme.
The ComSol software used for the FEM calculations was funded by the EPSRC under the grant no. EP/L001470/1. We thank V. Savona for help with the FDTD calculations.
\appendix
\section{Normalization of resonant states}
\label{App:IN}

We prove in this section that the spectral representation \Eq{ML4} leads to the RS normalization condition \Eq{normaliz} and further to Eq.\,(\ref{normk}). To do so, we consider an analytic continuation $\bE(k,\br)$ of the wave function $\En(\br)$ around the point $k=k_n$ in the complex $k$-plane ($k_n$ is the wavenumber of the given RS). We choose the analytic continuation such that it satisfies the outgoing wave boundary condition and Maxwell's wave equation
\be
- \nabla\times\nabla\times \bE(k,\br)+k^2\heps(\br)\bE(k,\br)=(k^2-k_n^2)\bs(\br)
 \label{MEcont}
\ee
with an arbitrary  source term corresponding to the current density ${\bf j}(\br)=\bs(\br) ic(k^2-k_n^2)/(4\pi k)$.
The source $\bs(\br)$ has to be zero outside the volume $V$ of the inhomogeneity of $\heps(\br)$ for the electric field $\bE(k,\br)$ to satisfy the outgoing wave boundary condition. It also has to be non-zero somewhere inside $V$, as otherwise $\bE(k,\br)$ would be identical to $\En(\br)$. We further require that $\bs(\br)$ is normalized according to
\be
\int_V\En(\br)\cdot\bs(\br)\,d\br=1+\delta_{k_n,0}\,,
\label{sigma}
\ee
with the Kronecker delta $\delta_{k_n,0}=1$ for $k_n=0$ and $\delta_{k_n,0}=0$ for $k_n\neq0$. This ensures that the analytic continuation reproduces $\En(\br)$ in the limit $k\to k_n$. Indeed, solving \Eq{MEcont} with the help of the GF and using the GF spectral representation \Eq{GFexpansion}, we find:
\bea
\bE(k,\br)&=&\int_V \GFk(\br,\br') (k^2-k_n^2)\bs(\br') d\br'
\\
&=&\sum_{n'}\Em(\br) \frac{k^2-k_n^2}{2k(k-k_n)} \int_V \Em(\br')\cdot\bs(\br') \,d\br'\,,\nonumber
\eea
and using \Eq{sigma} obtain
$$
\lim_{k\to {k_n}} \bE(k,\br)=\En(\br)\,.
$$

We now consider the integral
\be
I_n(k) =\frac{\int_V(\bE\cdot\nabla\times\nabla\times \En -\En\cdot\nabla\times\nabla\times \bE)d\br}{k^2-k_n^2}
\label{I0}
\ee
and evaluate it by using Maxwell's wave Eqs.\,(\ref{me3D}) and (\ref{MEcont}) for $\En$ and $\bE$, respectively, and the source
term normalization \Eq{sigma}:
\be
I_n(k) =\frac{\int_V (k_n^2\bE\cdot\heps \En -k^2\En\cdot\heps \bE)d\br}{k^2-k_n^2} +1+ \delta_{k_n,0}\,.
\label{I1} \ee
On the other hand, rearranging the integrand in \Eq{I0} and using the divergence theorem, we obtain
\be
(k^2-k_n^2)I_n(k) = \oint _{S_V} dS
\left(\En\cdot\frac{\partial\mathbf{E}}{\partial s}-\mathbf{E}\cdot\frac{\partial\En}{\partial
s}\right)
 \label{I2}
\ee
with $S_V$ being the the boundary of $V$.
Here, we used that for two arbitrary vector fields, $\ba(\br)$ and $\bb(\br)$, we can write
\bea
&&\ba\cdot\nabla\times\nabla\times \bb -\bb\cdot\nabla\times\nabla\times \ba =
\nonumber\\
&&\ba\cdot [\nabla (\nabla\cdot \bb)-\nabla^2\bb]-\bb\cdot [\nabla (\nabla\cdot \ba)-\nabla^2\ba] =
\nonumber\\
&&\nabla\cdot[\ba (\nabla\cdot \bb)-\bb (\nabla\cdot \ba)]+\hspace*{-3mm}\sum_{j={x,y,z}} \hspace*{-1mm}\nabla\cdot\left(-a_j\nabla b_j+b_j\nabla a_j\right )\,.
\nonumber
\eea
The divergence theorem therefore allows us to convert all volume integrals in \Eq{I0} into surface integrals over the closed surface $S_V$, the boundary of $V$, taken with an infinitesimal extension to the outside area where $\heps(\br)$ is homogeneous, so that both $\nabla \cdot \bE$ and $\nabla \cdot \En$ vanish on that surface leaving only the integral shown in \Eq{I2}.
Finally, using \Eq{I1} in \Eq{I2} and taking the limit $k\to k_n$ we obtain the normalization condition \Eq{normaliz}.

The limit in \Eq{normaliz} can be taken explicitly for any spherical surface~\cite{MuljarovEPL10}. In fact, outside the system, where $\heps(\br)=\hat{\mathbf{1}}$ (or a constant) the wave function of any $k_n\neq 0$ mode
is given by $ \En(\br)=\Fn(k_n \br)$, where $\Fn({\bf q})$ is a vector function satisfying the equation
\be
\nabla_{\bf q}\times\nabla_{\bf q}\times \Fn ({\bf q})=\Fn ({\bf q})
\ee
and the proper boundary conditions at system interfaces and at ${\bf q}\to \infty$. The analytic continuation of $\En(\br)$ can be therefore be taken in the form \be
\bE(k,\br)=\Fn(k \br)\,.
\ee
We use a Taylor expansion at $k=k_n$ to obtain
\bea
\bE(k,\br)&\approx &\Fn(k_n \br)+(k-k_n)r\left.\frac{\partial \Fn(k{\bf r})}{\partial (kr)}\right|_{k=k_n}
\nonumber\\
&=&\En(\br)+\frac{k-k_n}{k_n}r\,\frac{\partial \En(\br)}{\partial r}
\label{ETaylor}
\eea
and
\be
\frac{\partial \bE(k,\br)}{\partial r}\approx \frac{\partial \En(\br)}{\partial r}+\frac{k-k_n}{k_n}\frac{\partial }{\partial r}r\,\frac{\partial \En(\br)}{\partial r}\,,
\label{Eder}
\ee
where $r=|\br|$ is the radius in the spherical coordinates. Choosing the origin to coincide with the center of the sphere of integration $S_V=S_R$ we note that $\partial/\partial s= \partial/\partial r$  in \Eq{normaliz}. Substituting Eqs.\,(\ref{ETaylor}) and (\ref{Eder}) into \Eq{normaliz} and taking the limit $k\to k_n$ obtain \Eq{normk}.

\section{Matrix elements for various perturbations in 3D}
\label{App:ME}
In this section we give explicit expressions for the matrix elements $V_{nn'}$ calculated for the homogeneous perturbation treated in \Sec{sec:Hom} and for a perturbation in the form of a piece of a homogeneous spherical shell layer. The latter is suitable
for treating an arbitrary symmetric or asymmetric perturbation of the sphere and is used in particular for half- and quarter-sphere  perturbations considered in \Sec{sec:Halfmoon} and \ref{sec:quartermoon}, respectively.

\subsection{Homogeneous sphere perturbation}
\label{App:ME_TE}
The homogeneous perturbation \Eq{eps-hom} does not mix different $m$ or $l$ values, nor does it mix TE modes with TM or LE modes. Using the definition \Eq{Vnm} we calculate the matrix elements between TE RSs performing the angular integration which leads to the $lm$-orthogonality:
\be
V^{\rm TE}_{nn'}\!=\!\Delta\epsilon\, l(l+1)\delta_{ll'}\delta_{mm'} (A^{\rm TE}_l)^2 \!\!\int_0^R\!\!\! R_l(r,k_n)R_l(r,k_{n'}) r^2dr.
\nonumber\label{App:TE}
\ee
The radial integration can also be done analytically, so that the matrix elements take the form
\be V^{\rm TE}_{nn}= \frac{\Delta\epsilon}{n_R^2-1}\left[1-\frac{j_{l-1}(x)j_{l+1}(x)}{j^2_{l}(x)}\right]
\label{App:TE1}
\ee
for identical basis states $n=n'$ and
\be V^{\rm TE}_{nn'}= \frac{\Delta\epsilon}{n_R^2-1}\, \frac{2\delta_{ll'}\delta_{mm'}}{x^2-y^2}\left[\frac{y j_{l-1}(y)}{j_{l}(y)}-\frac{x j_{l-1}(x)}{j_{l}(x)}\right]
\label{App:TE2}
\ee
for different basis states $n \neq n'$, where $x=n_R k_{n} R$ and $y=n_R k_{n'} R$\,. Similarly, for TM RSs we find
\bea
\!\!\!\!\!\!\!\!\!\!V^{\rm TM}_{nn'}&=&\frac{\Delta\epsilon \,l(l+1)}{n_R^4 k_n k_{n'}}\delta_{ll'}\delta_{mm'} A^{\rm TM}_l (k_n) A^{\rm TM}_l (k_{n'})
\nonumber\\
&&\times
\int_0^R \Biggl\{l(l+1) R_l(r,k_n)R_l(r,k_{n'})
\nonumber\\
&&\ \ \ \ \ \ \ \ +\frac{\partial [r R_l(r,k_n)]}{\partial r}\frac{\partial [r R_l(r,k_{n'})]}{\partial r } \Biggr\} dr\, ,
\nonumber\label{App:TM}
\eea
and after analytic integration we obtain
\be V^{\rm TM}_{nn}= \frac{\Delta\epsilon}{n_R^2-1}\,\frac{1}{F_l(x)}\left[2\frac{l+1}{x^2}+\frac{j^2_{l+1}(x)}{j^2_{l}(x)}-\frac{j_{l+2}(x)}{j_{l}(x)}\right]
\label{App:TM1}
\ee
for identical basis states $n=n'$ and
\bea
V^{\rm TM}_{nn'}&=& \frac{\Delta\epsilon}{n_R^2-1}\,\frac{1}{\sqrt{F_l(x)F_l(y)}}\, \frac{2\delta_{ll'}\delta_{mm'}}{x^2-y^2}
\label{App:TM2}\\
&&\times \left[(l+1)\frac{x^2-y^2}{x y}+\frac{y j_{l+1}(x)}{j_{l}(x)}-\frac{x j_{l+1}(y)}{j_{l}(y)}\right]
\nonumber
\eea
for different basis states $n \neq n'$, where
\be
F_l(x)=\left[\frac{j_{l-1}(x)}{j_l(x)}-\frac{l}{x}\right]^2+\frac{n_R^2 l(l+1)}{x^2}\,,
\label{App:F}
\ee
with $x=n_R k_{n} R$ and $y=n_R k_{n'} R$\,. Note that LE and TM modes are mixed by the perturbation, and non-vanishing matrix elements between them are calculated using Eqs.\,(\ref{App:TM1}) and (\ref{App:TM2}),
treating the LE modes as TM modes with $k_n=0$ and the normalization constants multiplied by $\sqrt{l(n_R^2-1)}$, in agreement with \Eq{TM2TE}.

\subsection{Arbitrary perturbations}
\label{App:Arbpert}
An arbitrary perturbation of the sphere can be treated as a superposition of homogeneous perturbations in the form of  spherical-shell pieces, each piece described by
\be
\Delta \varepsilon({\bf r})= \left\{
\begin{array}{cl}
\Delta \epsilon & {\rm for}\ \
\begin{array}{c}
R_1\leqslant r\leqslant R_2 \\
\theta_1\leqslant \theta\leqslant \theta_2 \\
\varphi_1\leqslant \varphi\leqslant \varphi_2
\end{array}
\\
&\\
0 & {\rm otherwise}.\\
\end{array} \right.
\label{deleps}
\ee
The hemisphere perturbation \Eq{eps-half} is then described by \Eq{deleps} with
\mbox{$0\leqslantt r\leqslantt R$,} \mbox{$0\leqslantt \theta\leqslantt \pi/2$},  and $0\leqslantt\varphi\leqslantt 2\pi$\,. The quarter sphere perturbation \Eq{eps-Quarter} is given by \Eq{deleps} with
$0\leqslantt r\leqslantt R$, $0\leqslantt \theta\leqslantt \pi/2$, and $\pi/2\leqslantt \varphi\leqslantt 3\pi/2$\,.

Factorizing the radial and angular integrals and using the fact that $\chi'_m(\varphi)=m\chi_{-m}(\varphi)$, the matrix elements of the perturbation \Eq{deleps} become
\bea
V^{\rm TE}_{nn'}&=&\Delta \epsilon \,A^{\rm TE}_l A^{\rm TE}_{l'}
\label{App:TETE}\\
&&\times T_{1;nn'}^{ll'} \left(mm'S_{-m}^{-m'}Q_{1;ll'}^{mm'}+S_m^{m'}Q_{2;ll'}^{mm'}\right)
\nonumber
\eea
between TE modes,
\bea
V^{\rm TM}_{nn'}&=&\Delta \epsilon \,\frac{A^{\rm TM}_{l} (k_{n})A^{\rm TM}_{l'} (k_{n'})}{n_R^4 k_n k_{n'}}
\label{App:TMTM}\\
&&\times\left[ l(l+1)l'(l'+1) T_{2;nn'}^{ll'} S_{m}^{m'}Q_{3;ll'}^{mm'}\right.
\nonumber\\
&&\left.+T_{3;nn'}^{ll'} \left(mm'S_{-m}^{-m'}Q_{1;ll'}^{mm'}+S_m^{m'}Q_{2;ll'}^{mm'}\right)\right]
\nonumber
\eea
between TM modes, and
\bea
V^{\rm TE-TM}_{nn'}&=&\Delta \epsilon \,A^{\rm TE}_l \frac{A^{\rm TM}_{l'} (k_{n'})}{n_R^2 k_{n'}}
\label{App:TETM}\\
&&\times T_{4;nn'}^{ll'} \left(mS_{-m}^{m'}Q_{4;ll'}^{mm'}-m'S_m^{-m'}Q_{4;l'l}^{m'm}\right)
\nonumber
\eea
between TE and TM modes.
The integrals contributing to Eqs.\,(\ref{App:TETE}), (\ref{App:TMTM}), and (\ref{App:TETM}) are given by
\bea
T_{1;nn'}^{ll'}&=&\int_{R_1}^{R_2} \bar{j}_l(n_R k_n r)\bar{j}_{l'}(n_R k_{n'} r)r^2 dr\,,
\nonumber\\
T_{2;nn'}^{ll'}&=&\int_{R_1}^{R_2}  \bar{j}_l(n_R k_n r)\bar{j}_{l'}(n_R k_{n'} r)dr \,,
\nonumber\\
T_{3;nn'}^{ll'}&=&\int_{R_1}^{R_2} \frac{d}{dr}\Bigl[r\bar{j}_l(n_R k_n r)\Bigr]\frac{d}{dr}\Bigl[r\bar{j}_{l'}(n_R k_{n'} r)\Bigr]dr\,,
\nonumber\\
T_{4;nn'}^{ll'}&=&\int_{R_1}^{R_2} \bar{j}_l(n_R k_n r)\frac{d}{dr}\Bigl[r\bar{j}_{l'}(n_R k_{n'} r)\Bigr]r dr\,,
\nonumber\label{App:T}\\
S_{m}^{m'}&=&\int_{\varphi_1}^{\varphi_2} \chi_m(\varphi)\chi_{m'}(\varphi)d\varphi\,,
\nonumber\label{App:S}
\\
Q_{1;ll'}^{mm'}&=&\int_{\theta_1}^{\theta_2} \frac{\bar{P}_l^m(\cos\theta)\bar{P}_{l'}^{m'}(\cos\theta)}{\sin\theta}d\theta \,,
\nonumber\\
Q_{2;ll'}^{mm'}&=&\int_{\theta_1}^{\theta_2} \frac{d}{d\theta}\left[\bar{P}_l^m(\cos\theta)\right] \frac{ d}{d\theta}\left[\bar{P}_{l'}^{m'}(\cos\theta)\right]\sin\theta d\theta\,,
\nonumber\\
Q_{3;ll'}^{mm'}&=&\int_{\theta_1}^{\theta_2} \bar{P}_l^m(\cos\theta)\bar{P}_{l'}^{m'}(\cos\theta)\sin\theta d\theta\,,
\nonumber\\
Q_{4;ll'}^{mm'}&=&\int_{\theta_1}^{\theta_2} \bar{P}_l^m(\cos\theta)\frac{ d}{d\theta}\left[\bar{P}_{l'}^{m'}(\cos\theta)\right]d\theta\,,
\label{App:Q}
\eea
where
\be
\bar{j}_l(k r )\equiv\frac{{j}_l(k r)}{{j}_l(k R)}
 \ee
and
\be
\bar{P}_l^m(x)\equiv\sqrt{\frac{2l+1}{2}\frac{(l-|m|)!}{(l+|m|)!}}P_l^{|m|}(x)\,.
\label{App:P}
\ee

\end{document}